\documentclass[pre,aps,twocolumn]{revtex4}
\usepackage{graphicx}

\def\fk{$F(k)$ }
\def\aa{$\alpha$}

\begin{document}
\title{Critical Fluctuation of Wind Reversals in Convective
Turbulence}
\author{Rudolph C. Hwa$^1$, C.\ B.\ Yang$^{2}$,
S.\ Bershadskii$^3$, J.\ J.\ Niemela$^3$, and K.\ R.\ Sreenivasan$^3$}
\affiliation{$^1$Institute of Theoretical Science and Department of Physics\\
University
of Oregon, Eugene, OR 97403-5203, USA}
\affiliation{$^2$Institute of Particle Physics, Hua-Zhong Normal University, Wuhan
430079, P.\ R.\ China}
\affiliation{
$^3$International Center of Theoretical Physics, Strada Costiera 11, I-34100
Trieste, Italy}
\date{\today}

\begin{abstract}
The irregular reversals of wind direction in convective turbulence
are found to have fluctuating intervals that can be related to
critical behavior. It is shown that the net magnetization of a 2D
Ising lattice of finite size fluctuates in the same way.
Detrended fluctuation analysis of the wind reversal time series
results in a scaling behavior that agrees with that of the Ising
problem. The properties found suggest that the wind reversal
phenomenon exhibits signs of self-organized criticality.

\end{abstract}
\maketitle

In turbulent thermal convection at high Rayleigh numbers (Ra) it
has recently been observed that there exists not only large-scale
circulating motion, called mean wind, but also abrupt reversals of
the wind direction, whose physical origin is still largely unknown
\cite{kh}-\cite{ns2}. For our purposes, Ra is simply
a non-dimensional measure of the temperature difference between
the bottom and the top plates of the container within which the
convective motion occurs. Metastable states have been suggested to
describe the two opposite directions of the wind, and the reversal
of its direction is to be understood in terms of the imbalance
between buoyancy effects and friction \cite{sbn}. Instead of
searching for the origin  of the wind reversals in the framework
of hydrodynamical considerations, we investigate in this paper the
possibility of understanding the phenomenon in a totally different
context, namely: critical phenomenon.  We shall find a measure to
quantify the fluctuations in the wind direction, and then
demonstrate that its behavior corresponds to one exhibited by a
system undergoing a second-order phase transition.  We then
perform a detrended fluctuation analysis to determine the detailed
properties of the fluctuations of the wind, more specifically its
scaling behavior.

The experimental data that we analyze are the same as those
reported in \cite{nss} and studied in  \cite{sbn}. By varying the
pressure and lowering the temperature of the gas, the Rayleigh
number could be varied between $10^6$ and $10^{16}$. Further
details of the apparatus can be found in \cite{nss}. We focus on
the data that give the wind speed and direction for a continuous
period of up to one week at Ra$=1.5\times 10^{11}$. Figure 1 shows
a small segment of the wind velocity data for  6.5 hr, starting at an arbitrary time. Note
how the wind changes direction suddenly in the time scale of that
figure. We proceed directly to an interpretation of the
fluctuations of the wind velocity.

In a fully developed turbulent convection at high Ra there are two
opposing dynamical features. One is the emission of plumes from
the top and bottom boundary layers; they occur at random locations
and at random times in varying sizes.  The other is the existence
of mean wind that rotates in one direction or another, making
rapid reversals at seemingly random intervals.  We regard the
former as the disordered motion of the components of a complex
system, and the latter as the ordered motion of the whole of the
system.  For low Ra (say below $10^9$), the ordered motion is not
sufficiently impeded by the disordered motion to cause reversals
of the wind direction. At high Ra, the cumulative effect of the
many plumes that is strong enough to reverse the wind direction.
The system then proceeds as before except that the wind rotates in
the opposite direction with varying magnitude until another
reversal occurs due to the collective action of the disorganized
plumes. If the system is at a critical state, whether
self-organized  or not, the competition between the ordered and
disordered motions leads to the wind switching directions at
irregular intervals of all scale. The probability of occurrence of
the wind duration $\tau$ between reversals should satisfy a power
law
\begin{equation} p(\tau) \sim \tau^{-\gamma}     \label{1}
\end{equation} as a manifestation of criticality. Such a power
law has been found in the data \cite{sbn}. The discussion above
describes our view of the origin of such a scaling behavior.

We now advance the idea that the above description of the wind
and plumes in convective turbulence in terms of ordered and
disordered motions has its corresponding counterparts in the 2D
Ising model of critical behavior.  In the Ising system of
near-neighbor interactions without external magnetic field the
lattice spins tend to align in the same direction except for the
random disorientation due to thermal fluctuation.  For a finite
lattice the net magnetization, $M$, is non-vanishing. For
$T<T_c$, the critical temperature, $M$ is likely to persist in
the same direction for  longer time in lattice-spin updating than
at higher $T$. At $T>T_c$ the thermal interaction dominates, and
$M$ is more likely to flip sign more frequently upon updating. The
fluctuation of the signs of $M$ is therefore a property  that
reflects the tension between the ordered and disordered
interactions of the whole system.

\begin{figure}[tbph]
\includegraphics[width=0.45\textwidth]{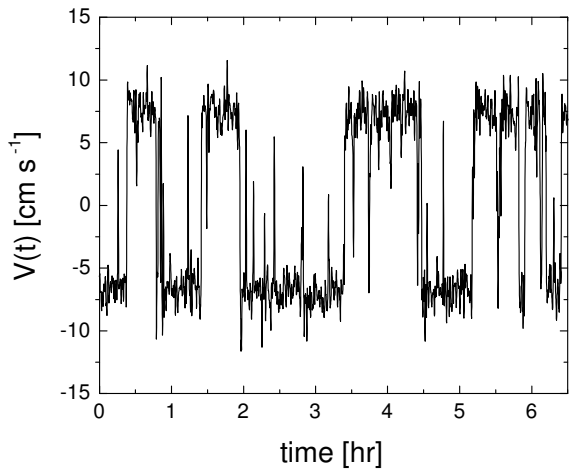}
\caption{ A segment of the data on wind
velocity fluctuation at Ra$=1.5\times 10^{11}$. The data were obtained
from an apparatus that was essentially a
cylindrical container of 50 cm diameter and 50 cm height, filled with
cryogenic helium gas. Two sensors were placed outside the boundary layer
on the sidewall of the container near the middle section of the
container, and were separated vertically by a distance of
1.27 cm.}
\end{figure}

Since the mean wind is a global phenomenon in a vessel of finite
volume, it is sensible for us to associate the wind direction
with the sign of $M$ of the Ising lattice of finite size. We can
then map wind reversal to the reversal of $M$ upon updating the
lattice spins in a simulation. The plumes are the disordered
fluctuations that correspond to the spin fluctuations due to
thermal agitation, and the wind is the ordered motion that can
change direction just as the magnetization can change sign when
enough lattice spins change directions. The key connection
between the two problems is the mapping of the real time in
turbulence to the time of updating the Ising configurations. It
is therefore crucial that each configuration has some memory of
the previous configuration before updating; hence we employ the
Metropolis algorithm, which does precisely this. It should be
noted that we are entering into a rather unexplored territory
where the process of computer simulation itself is endowed with
some physical significance, quite unrelated to the large body of
analytical work that has been devoted to the Ising model of
infinite lattice. Our task is to show that the wind reversal
problem corresponds to the Ising problem of finite lattice at the critical
temperature.

To be more specific, we consider a square lattice of size $L^2$,
where $L$ is taken to be 255, an odd number. We start with the
$L^2$ site spins having a random distribution of $\pm 1$ values.
We then visit each site and determine from the usual near-neighbor
interaction whether its spin should be reversed: yes, if the
energy is lowered by the flip; if not lowered, the flip can still
take place according to a thermal distribution specified by
temperature $T$. One time step is taken by the whole system when
all sites are updated. We take $3 \times 10^5$ time steps in
total, and divide the whole series into 30 segments. The values of
$M$ at each of the $10^4$ time points in each segment are
discretized to $\pm 1$, according to $M{}^{>}_{<}0$. A continuous
string of $M$ of one sign, either +1 or $-1$, forms a duration
that is analogous to the mean wind rotating in one direction. The
reversals of $M$ correspond to the reversals of wind. Near the
critical point, durations of all lengths can occur.

Before considering the issue of criticality for a finite lattice,
let us discuss the measure that we shall use for quantifiying the
duration fluctuations appropriate for both the wind and Ising
problems. The experimental data on wind consist of 8 segments,
each having ${\cal T}=10,282$ time points. For the Ising case we
have $30$ segments, each having ${\cal T}=10^4$, roughly the same
as wind data. Let $N$ denote the number of reversals in a segment.
With the locations of the reversals denoted by $t_i$, $i = 1$,
$\cdots$, $N$, define $\tau_i = t_{i+1}-t_i$ to be  the $i$th
duration (or gap), where $t_0$ and $t_{N+1}$ are assigned to be
the left and right ends of the segment, respectively.  Now, define
the moment \cite{hz1}
\begin{eqnarray}  G_q = {1  \over  N+1} \sum^N_{i = 0}\left({ \tau _i \over
{\cal T} } \right)^q ,
\label{2}
\end{eqnarray}  where $q$ is any positive integer.  Clearly, $G_0 = 1$ and
$G_1 = 1/(N+1)$. $G_q$ is a measure that quantifies the pattern of
reversals in each segment. For large $q, G_q$ is a small number,
since $\tau_i/{\cal T}$ is small. Its value can be dominated by a
few large gaps, as when $T<T_c$, or may become
significant from the sum over many small contributions due to many
small gaps, as when $T>T_c$. For a measure of the fluctuations of
$G_q$ from segment to segment, we define an entropy-like quantity
\cite{hz1,hz2}
\begin{eqnarray}  S_q = - \left< G_q  \ln G_q \right> ,
\label{3}
\end{eqnarray}  where $\left<\cdots\right>$ implies an average over all
segments.  For brevity we shall refer to the study of the time
series in terms of $S_q$  as the gap analysis.  In Fig.\ 2 we show
by filled circles the result of the gap analysis on the wind data
at Ra$=1.5\times10^{11}$. It is evident that for $q\ge 2$ the
points can be well fitted by a straight line, shown by the solid
line, exhibiting an exponential behavior for $S_q$
\begin{eqnarray}
\ln S_q = - \lambda q + \lambda_0 \ ,  \quad\quad   \lambda=0.264\ .
\label{4}
\end{eqnarray}

For the Ising simulation we must first decide on the proper value
of the critical temperature $T_c$ for a finite lattice. For an
infinite lattice its value has been determined analytically to be
2.269 in units of $J/k_B$, where $J$ is the coupling strength of
near-neighbor interaction and $k_B$ the Boltzmann constant
\cite{khu}. For a finite lattice the value of $T_c$ should be
higher.  We have performed the simulation of our Ising system at
three values of $T$, and determined the properties of $M$
reversal. In Fig.\ 2 we show the results of our calculated values
of $S_q$ at $T=2.305, 2.310$ and 2.315. Only the one at $T=2.310$
(lowered by a factor of 2 for clarity) shows a nearly linear
dependence in the plot. The dashed line is a linear fit of the
points in open circle, giving a slope of $\lambda=0.261$. At the
two neighboring values of $T$, the $q$ dependencies of log$S_q$
(shown by triangles and squares) are not linear, the values at
high $q$ being higher than at $T=2.310$. The linear behavior at
$T=2.310$ is almost the same as in the wind reversal problem, as
can be seen visually by the dash-dot line, which is a parallel
transport of the solid line for comparison, but displaced slightly
from the dashed line to avoid overlap. We regard $T=2.31$ as the
critical temperature $T_c$ in our Ising system, since it has the
unique property of being different from those of the neighboring
$T$ on both sides.  When $T<T_c$, the gaps are longer and $G_q$ is
larger at large $q$ (but still $\ll 1$) with the consequence that
$S_q$ is larger. When $T>T_c$, the gaps are shorter, but many gaps
can contribute in the sum in Eq.\ (\ref{2}), resulting in $G_q$
still being larger at large $q$ with the consequence that $S_q$ is
also larger. It is only at the critical point that gaps of all
sizes can occur, resulting in $G_q$ to be smaller and therefore
$S_q$ also smaller at large $q$. Thus the exponential decrease of
$S_q$ is a signature of criticality. The value of $T_c$ obtained
here is in accord with the result of another
calculation, in which the normalized factorial moments are found
to exhibit scaling behavior at the critical point, different from
the non-scaling behaviors at neighboring $T$ \cite{cghwa}. In that
calculation the measure studied quantifies the fluctuation of the
cluster sizes in an Ising system on a square lattice of size
$L=288$, for which $T_c$ is found to be 2.315.

\begin{figure}[tbph]
\includegraphics[width=0.45\textwidth]{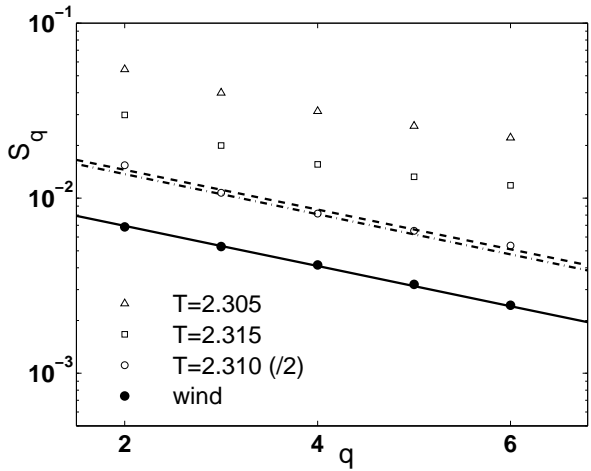}
\caption{ Moments in the gap analysis of wind reversal (filled
circles) and magnetization reversal in Ising lattice (open symbols)
for different temperatures. The open circles are lowered by a factor
of 2 to give space for clarity. The solid line is a linear fit of
filled circles, and the dashed line is a linear fit of open circles.
The dash-dot line is parallel to the solid line, placed near the
dashed line for comparison.}
\end{figure}

The normalizations of $S_q$ for the wind and Ising problems are
not the same, since the average numbers $N$ of reversals are
different. However, the exponential behaviors are remarkably
identical. The $q$ dependence of $S_q$ is a quantitative measure
of the fluctuation behavior of the reversals. The fact that the
slope $\lambda$ is the same for both the wind and magnetization
problems suggests strongly that the wind reversal in convective
turbulence at high Ra is a critical phenomenon. Moreover, since we
have not tuned any adjustable parameter in the wind problem to
bring the system to the critical point, as we have done for the
Ising system by varying $T$, we conclude that the wind reversal
phenomenon is a manifestation of self-organized criticality (S0C)
\cite{pb}.

We now search for a power-law behavior that characterizes changes
in the wind direction. (For other such efforts, see \cite{sbn}).
Our method is the detrended fluctuation analysis (DFA), which has
been found to reveal the scale-independent nature of time series
in a variety of problems, ranging from heartbeat irregularity
\cite{ckp} and EEG \cite{hf} to economics \cite{hu}. In that
analysis we look for scaling behavior in the RMS deviation of the
wind velocity from local linear trends. Given the time series of
the wind velocity $V(t)$ over a total range of ${\cal T}_{\rm
max}$, we divide it into $B$ equal bins of width $k$, discarding
the remainder ${\cal T}_{\rm max}-Bk$. Let $\bar V_b(t)$ denote
the linear fit of $V(t)$ in the $b$th bin. The variance of the
deviation of $V(t)$ from the local trend, $\bar V_b(t)$, in bins
of size $k$ is defined by
\begin{equation} F^2(k)={1\over B}\sum_{b=1}^B {1\over k}
\sum_{t=t_1}^{t_2} [V(t)-\bar V_b(t)]^2 \ ,   \label{6}
\end{equation}
where $t_1=1+(b-1)k$ and $t_2=bk$, measured in units of $\Delta
t=5 $ sec, so that the values of $t$ are dimensionless integers
that count the time points in the data. The goal is to study the
behavior of the RMS fluctuations $F(k)$, as $k$ is varied. If
there is no characteristic scale in the problem, then $F(k)$
should have a scaling behavior
\begin{equation} F(k) \propto k^\alpha\ .     \label{7}
\end{equation} This power law cannot be valid for arbitrarily large $k$
because the series $V(t)$ is bounded, so for very large $k$ the linear trend
is just the $V(t)=0$ line, and the RMS fluctuation $F(k)$ must become
independent of $k$. Thus we expect ln$F(k)$ to saturate and deviate from
(\ref{7}) at some large $k$. We note parenthetically that we have applied
DFA to the unintegrated time series $V(t)$, which is a departure from the
usual practice.

\begin{figure}[tbph]
\includegraphics[width=0.45\textwidth]{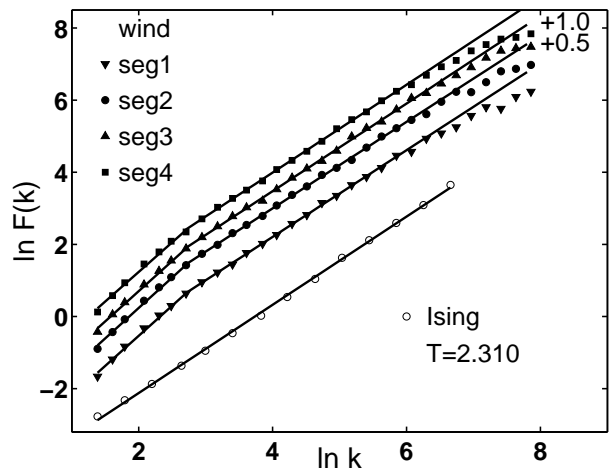}
\caption{Scaling behaviors of $F(k)$ in DFA of wind reversal
(filled symbols) and magnetization reversal in Ising lattice at the
critical temperature (open circles). Lines are linear fits.}
\end{figure}

In Fig.\ 3 we show \fk in a log-log plot for four equal segments
of the complete wind data in solid symbols. The segment seg1 is
for time running from 0 to 116,435 s, corresponding to ${\cal
T}_{\rm max}=23,287$; other segments all have the same length. We
have limited the maximum bin size to 2,580, so that even for the
largest bin the fluctuations can be averaged over 9 bins.
Evidently, there is a good scaling for each segment. The points
for seg3 and seg4 are shifted upwards by the quantities indicated
in order to give clarity without overlap. Note that the seg1 data
do not have the same magnitude of \fk as the other segments; yet
the scaling exponents are essentially the same. The deviation from
the straight lines at the upper end is the saturation effect
already discussed. There is another short region of scaling with a
higher slope at low $k$. It is a consequence of fluctuations of
the velocity within one direction of the wind, whose presence is
evident in Fig.\ 1. Since the critical behavior identified here
refers to wind reversals, and not to fluctuations of the wind
velocity within one direction, we should ignore the lower short
scaling region.

In the scaling region to which we pay attention here, the slopes
are \aa=1.20, 1.20, 1.21 and 1.22, for seg1 to seg4, respectively.
The deviations among the segments are obviously small. The average
value is
\begin{equation}
  \alpha=1.21.    \label{8}
\end{equation}  This large value of \aa\  implies a smoother
landscape compared to the rough time series of white noise that is
characterized by complete unpredictability \cite{ckp}. Indeed, the
fluctuations of the wind reversal time series has gaps of all
sizes, the signature of critical behavior that is characterized by
$1/f$ noise \cite{pb}. It is interesting to compare our result
with the properties of the power spectral density for the velocity
found in Ref.\ \cite{nssd}, where a scaling behavior is shown to
exist with a slope roughly -7/5 (not by fitting) in the region
$-3<\log f<-1.8$. That range of frequency corresponds to
$4.1<\ln(1/f)<6.9$. If we identify the values of $k$ in DFA to the
time scale $1/f$, then that range of $\ln(1/f)$ corresponds  to
the range of $\ln k$ in Fig.\ 3, in which we find the scaling
behavior with the exponent \aa\ given in Eq.\ (\ref{8}). That
value of \aa\ is not too different from 7/5. The scaling behavior
found in DFA uses shorter segments of the whole data and exhibits
the power law more precisely, from which the value of \aa\ can be
more accurately determined.

We now apply DFA to the Ising problem. We consider 10 segments of
the $M$ reversal time series of the Ising lattice set at $T_c$,
each segment having $10^4$ time points. From the $F(k)$
determined in each segment, we average over all segments and show
the resultant dependence on $k$ in Fig.\ 3 by the open circles.
Clearly, the points can be well fitted by a straight line. The
slope is
\begin{eqnarray}
\alpha_M=1.22,   \label{9}
\end{eqnarray}
which is essentially the same as that in Eq.\ (\ref{8}) for wind
reversal. With the equivalence of these two scaling behaviors
established, we have found stronger evidence that the wind
reversal problem is a critical phenomenon.

To summarize, we have studied the time series of wind reversal in
convective turbulence by two methods (gap analysis and detrended
fluctuation analysis) and applied the same methods to the time
series of the reversal of the net magnetization of a 2D Ising
lattice. The same results are obtained for both problems. We
therefore can assert that wind reversal exhibits all the essential
properties characteristic of a critical behavior; apparently
requiring no tuning, it can be regarded as self-organized.

This work was supported, in part,  by the U.\ S.\ Department of
Energy under Grant No. DE-FG03-96ER40972, and by the National
Science Foundation under Grant No. DMR-95-29609.

\end{document}